# Selective excitation of homogeneous spectral lines

## A. K. Khitrin

*Department of Chemistry, Kent State University, Kent, OH 44242*

**Abstract**

It is possible, for homogeneously broadened lines, to excite selectively the response signals, which are orders of magnitude narrower than the original lines. The new type of echo, which allows detecting such signals, and the formalism, useful for understanding the phenomenon, as well as the experimental examples from NMR spectroscopy are presented.

## I. INTRODUCTION

In Fourier-transform spectroscopy, the signals are produced by excitation pulses which can be either "hard", when interaction with the pulse is stronger than internal interactions, or "soft" selective pulses, when interaction with the pulse is weak compared to the Hamiltonian of internal interactions. Action of a "hard" pulse can often be described as a simple rotation of the initial density matrix. Excitation by a "soft" pulse involves more complex dynamics. It is convenient to discriminate between homogeneous and inhomogeneous line broadening. Inhomogeneous broadening comes from a distribution of resonance frequencies of individual systems. Homogeneous broadening can be a result of relaxation or interactions between individual systems. The later is the main source of line-broadening in solid-state NMR, where individual spins are coupled by dipole-dipole interactions. Experimentally, the two types of broadening can be distinguished by performing a hole-burning experiment. Long and weak excitation pulse can produce a narrow "hole" in inhomogeneous line by selectively saturating a narrow region of the spectrum. The result for a homogeneous line is different: the intensity of the entire line is degraded homogeneously. Another difference is that inhomogeneous broadening can be exactly compensated by a single $\pi$-pulse (Hahn echo [1]), while reducing a width of homogeneously broadened line is a much more challenging task. Based on a failure of hole-burning, one may also conclude that, for homogeneous lines, it is impossible to excite a narrow response signal. However, such conclusion would be wrong. We will show below that selective excitation can produce signals, which are orders of magnitude narrower than the original homogeneous lines. Such signals can be useful, as an example, for NMR imaging of solid samples.

A simple model with inhomogeneous broadening is an ensemble of $N$ non-interacting two-level systems (spins) with distributed resonance frequencies. In this case, the effect of irradiation pulse can be easily calculated, since it involves evolution of independent two-level systems. For a cluster of $N$ dipolar-coupled spins ½ with $2^N$ energy levels, the number of allowed transitions is almost $2^{2N}$. For large $N$, even when the range of excited frequencies is very narrow (long excitation pulse with small amplitude) there are numerous near-resonance transitions between a given state and other states of the entire cluster. The result can be viewed as a percolation, when the population changes propagate through all quantum states via a network of allowed transitions. Irradiation produces two effects. First, it drives pairs of populations for each active transition towards common values. Second, it drives populations to instant quasi-equilibrium, because of the mentioned above connectivity of states. Such behavior is well described by the Provotorov's saturation theory [2,3] which provides coupled kinetic equations for Zeeman and dipolar spin temperatures (extension of this approach to the case of low spin temperatures is given in [4]).

Some time ago, we have found [5-8] that long and weak excitation pulses can produce extremely sharp response signals in many substances with dipolar-broadened NMR lines. It has been also demonstrated that such long-lived coherent response signals can be efficiently used for high-resolution MRI of solids [9,10] or studying diffusion in liquid crystals [11]. Qualitatively, generation of such signals was explained as a result of partial saturation [7]. However, at that time we failed to provide any quantitative model capable of predicting signal intensities and achievable widths. It has been found empirically that for real solids many factors, such as structure and network of spin-spin couplings, types of molecular motion, spin-lattice relaxation, all play their role. In most cases we studied, the largest intensity of the signal at a given pulse amplitude has been reached at a pulse width corresponding to a total flip angle $\varphi \approx 4\text{-}6\pi$. Clearly nonlinear character of the excitation process made a search of an adequate theoretical model even more challenging.

It seems that for small flip angles $\varphi < 1$ of the soft excitation pulse the theoretical analysis simplifies. As a result, some understanding of the complicated process of interaction between the excitation pulse and the system of interacting spins can be achieved.

## II. LINEAR RESPONCE

We assume that the initial density matrix is $S_Z$, and that the excitation $y$-pulse produces observable magnetization along the $x$-axis of the rotating frame. Conventional lineshape $I_0(\omega)$ is the Fourier transform of the free induction decay signal $M(t)$:

$$I_0(\omega) = \int dt \, exp(-i\omega t) \, M(t), \quad M(t) = Tr\{S_X \, S_X(t)\}, \quad S_X(t) = exp(-iHt) \, S_X \, exp(iHt), \quad (1)$$

where $H$ is the Hamiltonian of the system. Here and below the integration symbol without limits means integration between $-\infty$ and $\infty$. It is convenient to introduce the Fourier components of the operator $S_X(t)$ as

$$S_\omega = \int dt \, exp(-i\omega t) \, S_X(t), \quad (2)$$

so that

$$exp(-iHt) \, S_\omega \, exp(iHt) \, = S_\omega \, exp(i\omega t), \text{ and } S_X(t) = \int d\omega \, S_\omega \, exp(i\omega t). \quad (3)$$

Let us find the result of a weak pulse with the amplitude $\omega_1(t)$. For simplicity, we assume that the pulse is symmetric: $\omega_1(t) = \omega_1(-t)$ and $\omega_1(t) = 0$ at $|t| \geq T$, where $2T$ is the pulse duration. Weakness of the pulse will mean that the total flip angle $\varphi = \int_{-T}^{T} dt \, \omega_1(t) << 1$, i.e. it produces small effect on the total z-magnetization $S_Z$. The induced transverse signal at the end of the pulse can be written as

$$S_X(T) = \int d\omega \int_{-T}^{T} dt \, \omega_1(t) \, S_\omega \, exp(i\omega(T-t)). \quad (4)$$

This expression means that the transverse signal $S_X$, created at the moment $t$ with the rate $\omega_1(t)$ has been later dephased by the system's Hamiltonian during the time $T$-$t$. Since $\omega_1(t) = 0$ at $|t| \geq T$, the integration limits can be extended, and Eq.(4) rewritten as

$$S_X(T) = \int d\omega \ exp(i\omega T) \ S_\omega \int dt \ \omega_1(t) \ exp(-i\omega t) = \int d\omega \ exp(i\omega T) \ S_\omega \ C(\omega), \qquad (5)$$

where $C(\omega)$ is the Fourier transform of the pulse shape $\omega_1(t)$. Therefore, for long pulses, the spectral components $S_\omega$ are excited only within a narrow spectral range $C(\omega)$, which coincides with the spectrum of the excitation pulse. Let us now calculate the observable magnetization

$$M(T) = Tr\{S_X \ S_X(T)\} = Tr \int d\omega' \ S_{\omega'} \int d\omega \ exp(i\omega T) \ S_\omega \ C(\omega)$$

$$= \int d\omega \ exp(i\omega T) \ C(\omega) \ I_0(\omega), \qquad (6)$$

where $I_0(\omega) = Tr(S_{-\omega} \ S_\omega)$ is the conventional lineshape. Here we used the fact that $Tr(S_{\omega'} S_\omega)$ is not zero only for $\omega' = -\omega$. With the assumption that $C(\omega)$ is much narrower than $I_0(\omega)$, we will find that

$$M(T) = I_0(0) \int d\omega \ exp(i\omega T) \ C(\omega) = I_0(0) \ \omega_1(T) = 0. \qquad (7)$$

Since the pulse boundaries have been set arbitrarily, the signal also remains zero at any later moments of time. In other words, the signal with narrow spectral range $C(\omega)$ exists during the pulse, as it is expected from the linear-response theory [12], but it becomes perfectly dephased immediately after the end of the excitation pulse. Eq. (7) states that no linear response signal is excited by a pulse, whose spectrum is much narrower than the conventional spectrum of the system. The result looks surprising because, until now, we made no assumption about the Hamiltonian of the system. Therefore, it should also be valid for inhomogeneously broadened lines. At the same time, our experience tells us that selective excitation is possible for inhomogeneous lines. A well known example is slice-selection pulses in MRI.

For inhomogeneously broadened spectra, dephasing of individual spectral components (phase factors $exp(i\omega T)$ in Eq.(5)) can be reversed by applying hard $\pi_X$-pulse after the excitation pulse (Hahn echo). Alternatively, in the interaction frame, various types of echo, like Hahn echo [1] or "magic echo" [13,14] can be viewed as resulting from change of a sign of the system's Hamiltonian (time reversal). For inhomogeneous line, $\pi_X$ rotation changes the sign of the Hamiltonian $H$, but does not change $S_X$. Then, by using the definition of the spectral components $S_\omega$, we will find that the result of the refocusing pulse at $t = T$ on the components $S_\omega(T) = exp(i\omega T) \ S_\omega$ in Eq.(6) will be $S_\omega(T) \rightarrow \pi_X \rightarrow S_\omega^*(T) = exp(-i\omega T) \ S_\omega$. Subsequent evolution during the time $T$ will add phase factors $exp(i\omega T)$ and, at time $t = 2T$, the signal will be totally refocused. As an example, for a rectangular excitation pulse, the signal $M(t)$ will last from $t = T$ to $t = 3T$, and will be also rectangular. Replacement of the $\pi_X$ refocusing pulse by $\pi_Y$ will not change anything, except it also changes the sign of $S_X$ (see Eqs. (1) and (2)): $S_\omega(T) \rightarrow \pi_Y \rightarrow -S_\omega^*(T)$, so that the signal will be exactly the same but negative.

Fig. 1a shows excitation by a rectangular soft pulse, followed by hard $\pi_X$-pulse for inhomogeneously broadened spectrum. The sample is 1% $H_2O$ in $D_2O$. Proton spectrum is artificially broadened to 3.3 kHz by external $x$-gradient. Excitation is linear at small flip angles $\varphi$. The shape of the excited spectra coincides with the Fourier transform of the excitation pulse, although it becomes distorted at large flip angles $\varphi > 1$. The result of excitation without the refocusing $\pi$-pulse is shown in Fig. 1b. Clearly, the excitation is non-linear in $\varphi$: at small $\varphi$ the signal is absent, in accordance with Eq. (7). Besides that, the signals are broader than the spectrum of the excitation pulse, and they are negative. In this case, the free induction decay starts at zero and goes negative: first, gradually increasing its absolute amplitude, then, decreasing. Therefore, the total intensity of the excited spectrum is zero.

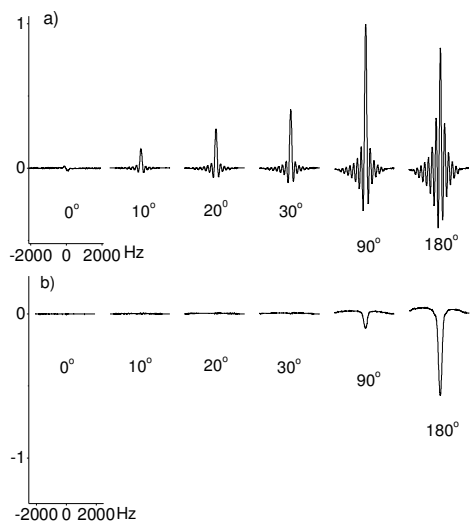

FIG. 1. *The signals in 1% $H_2O/D_2O$ sample for different flip angles of 4 ms rectangular excitation pulse, a) with and b) without $\pi_X$ refocusing hard pulse. The conventional spectrum is broadened to 3.3 kHz by external x-gradient.*

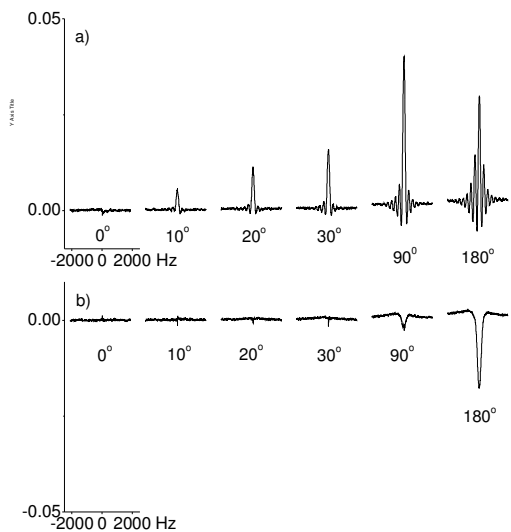

FIG. 2. *The signals in adamantane for different flip angles of 4 ms rectangular excitation pulse, a) with and b) without $\pi_X$ refocusing hard pulse. The amplitudes are in fractions of the conventional spectrum height.*

The refocusing pulses should not necessary be hard pulses. Soft pulses may act almost the same way, except that they truncate the "wings" of the spectra (oscillations for the rectangular excitation pulse). This gives an idea of qualitative explanation of excitation by a single soft pulse (Fig. 1b). Initial part of the pulse excites the spectral components, like in Eq. (5), while the rest of the pulse acts as a refocusing y-pulse, producing a negative signal. The overall result can be viewed as a superposition of pulses of different excitation lengths, which explains why the spectra are broadened.

Fig. 2 shows the result of the same experiment for solid adamantane. Adamantane is a plastic crystal with near-spherical fast-rotating molecules. Intramolecular dipole-dipole interactions between 16 protons of the same molecule are averaged by fast isotropic molecular rotations. Therefore, each proton spin is coupled to very large number of neighbors with close values of dipole-dipole constants. This is a perfect model system with homogeneous NMR line of about 12 kHz width. The results in Fig. 2 are almost identical to those in Fig. 1, except that the signal amplitudes are much smaller (the amplitudes are shown as fractions of the height of the conventional spectra, excited by a hard $\pi/2$-pulse). Oscillations in the spectra in Fig. 2a are slightly suppressed (the induction signals are less sharp steps) compared to the case of inhomogeneous broadening. The reason for this will be explained later. Similar to the case of inhomogeneous broadening, as a function of the flip angle of the refocusing pulse, the signal amplitude is maximal for π-pulses.

It is hard to imagine such long-lived response signals, surviving hundreds of almost random revolutions of individual spins in their local fields. However, for deterministic Hamiltonian dynamics, we would expect such collective response signals if there exists a way to spoil the perfect dephasing of the spectral components in Eq. (5). But here comes the problem. Experimentally, the signal reaches maximum for the refocusing π-pulse. Both the dipolar interactions and $S_X$ are invariant under $π_X$-rotation. Therefore, from Eqs. (1) and (2) we conclude that the refocusing pulse produces no effect on the spectral components, and no echo signal is expected. It is also easy to see that uniform rotation around $z$-axis cannot help. Therefore, a shift of resonance frequency (chemical shift) or macroscopic field inhomogeneity cannot produce an echo. However, the difference of chemical shifts can. The difference of chemical shifts does not commute with the Hamiltonian of dipole-dipole interactions, and changes its sign under $π_X$-rotation. There are two non-equivalent types of protons in adamantane, with the difference of isotropic chemical shifts about 0.1 ppm (50 Hz in our experiment). As some experimental support of the idea that the difference of chemical shifts is responsible for the echo formation, we can present an observation that the long-lived signal is undetectable in hexamethylbenzene, a plastic crystal with equivalent protons.

## III. THE ECHO MECHANISM

Let $H$ be the initial Hamiltonian and $H'$ is the Hamiltonian in the interaction frame (the coordinate system which rotates with the pulse) after the refocusing pulse at $t=T$. Similar to $H$, we can introduce the spectral operators $S_ω'$ for $H'$. Their evolution is simple: $S_ω'(t) = S_ω' exp(iωt)$, and the initial values can be calculated from projections on the density matrix in Eq.(5) by using the definitions in Eqs. (2) and (1). Calculation of the signal (magnetization) at times $t > T$ is straightforward, and the result is

$$M(t > T) = \int d\tau \int d\tau' \, G(\tau,\tau') \, ω_1(T - \tau) \, M_0'(\tau + \tau' + t - T), \qquad (8)$$

where $M_0'(t)$ is the conventional free-induction decay signal with the Hamiltonian $H'$, $ω_1(t)$ is the amplitude of the excitation pulse, and the correlator $G(\tau,\tau')$ is

$$G(\tau,\tau') = Tr\{ e^{iH'\tau'} e^{-iH\tau} S_x e^{iH\tau} e^{-iH'\tau'} e^{-iH'\tau'} S_x e^{iH'\tau'} \}. \qquad (9)$$

In this equation, the first "sandwiched" $S_X$ contains forward evolution with the Hamiltonian $H$ and backward evolution with $H'$, the second part is the forward evolution with the Hamiltonian $H'$. Eq. (8) shows that, for the echo signal, centered at $t = 2T$, to be observed, the correlator in Eq. (9) should have a long-lived component at $τ' = -2τ$. Along this direction, $G(\tau,\tau')$ can be written as

$$G(\tau,-2\tau) = Tr\{ e^{-iH'\tau} e^{-iH\tau} S_x e^{iH\tau} e^{iH'\tau} S_x \}. \qquad (10)$$

For the $π_X$ refocusing pulse, Eq. (10) can be also rewritten as

$$G(\tau,-2\tau) = Tr\{ e^{-iH\tau} e^{-i\pi S_x} e^{-iH\tau} S_x e^{iH\tau} e^{i\pi S_x} e^{iH\tau} S_x \}, \qquad (11)$$

which is exactly the expression for the echo amplitude in the Hahn echo experiment: $(π/2)_Y$-τ-$π_X$-τ, implemented by two hard pulses. Therefore, the selective excitation itself is not enough, and when sharp selective response signals exist, there should be also a long-lived Hahn echo. We will show below that such echo does really exist in systems with dipolar-broadened spectra.

Difference of chemical shifts is averaged by flip-flops of the spins. In other words, its Hamiltonian $\Delta$ does not commute with the Hamiltonian of dipole-dipole interactions $H_\mathrm{d}$, and gives only second-order corrections to the energy levels. Second-order corrections do not depend on the sign of the echo. Therefore, the mechanism of the echo is not obvious.

For two groups of states with magnetic quantum numbers $m$ and $m + 1$ the block structure of the operators $H_\mathrm{d}$, $\Delta$, and $S_\mathrm{X}$ is $\begin{pmatrix} H_d^1 & 0 \\ 0 & H_d^2 \end{pmatrix}$, $\begin{pmatrix} \Delta^1 & 0 \\ 0 & \Delta^2 \end{pmatrix}$, and $\begin{pmatrix} 0 & S_X \\ S_X & 0 \end{pmatrix}$. Therefore, we can introduce a simple three-level model with the Hamiltonian $H_\mathrm{d} + \Delta$ and the observable $S_\mathrm{X}$:

$$H = \begin{pmatrix} \omega_d & \Delta & 0 \\ \Delta & -\omega_d & 0 \\ 0 & 0 & 0 \end{pmatrix}, \qquad S_X = \begin{pmatrix} 0 & 0 & 1 \\ 0 & 0 & 1 \\ 1 & 1 & 0 \end{pmatrix}, \tag{12}$$

where a distribution of dipolar frequencies $\omega_\mathrm{d}$ models "pure" dipolar lineshape without contribution from chemical shift difference, and $\Delta$ represents the magnitude of the chemical shift difference. For the Hahn sequence $(\pi/2)_\mathrm{Y}$-$T$-$\pi_\mathrm{X}$, $S_\mathrm{X}$ evolves with the Hamiltonian $H$ during time $T$, then, at $t = T$, $\Delta$ changes the sign. This simple model predicts the echo at $t = 2T$ with the amplitude $M_0(\Delta/\Omega)$, where $M_0$ is the equilibrium magnetization and $\Omega$ is the width of distribution for $\omega_\mathrm{d}$. This echo amplitude is consistent with our experimental observations.

Another prediction of this three-level model is a peculiar behavior of the conventional spectrum near its center. At $\omega \to \Delta$, $\omega > \Delta$, the line diverges as $\omega/(\omega^2 - \Delta^2)^{1/2}$, and the line intensity is zero for $-\Delta < \omega < \Delta$. When interpreting this result we should note that $\Delta$ in Eq. (12) is not the difference between two specific chemical shifts, but the matrix element of the operator between two levels of the dipolar Hamiltonian. Therefore, it would be natural to assume that $\Delta$ is also distributed. The region of increased curvature at the line center increases the second moment of the free induction signal $\int dt M(t) t^2$ and makes it easier to observe the signals with long acquisition delays [15]. However, compared to the free induction signal, the echo signal is much stronger at long times. As an example, for adamantane the free induction signal after 1 ms delay is three orders of magnitude weaker than the echo signal, and it becomes practically undetectable beyond 1.5 ms, while the echo signal has the decay time about 13 ms. (It has been noticed earlier [16] that, for acquisition-delayed signals, introduction of the $\pi$-pulse increases the signal and makes it possible using longer delays.)

A more abstract and schematic explanation of the echo mechanism is given in Fig. 3. Each Hermitian spectral component $S_\omega + S_{-\omega}$ rotates in its own two-dimensional space of operators $S_\omega + S_{-\omega}$ and $i(S_\omega - S_{-\omega})$. At $t = T$ the "old" operators project on "new" operators as in Fig. 3a. Totally dephased components in Fig. 3b after projection change their lengths (and phases), as shown by dashed ellipse. This causes a partial echo at $t = 2T$ as in Fig. 3c.

From a fundamental point of view, the echo is a result of broken symmetry. The origin of the phenomenon is similar to liquid-state NMR experiments [17,18] where difference in chemical shifts has been used to access long-lived two-spin singlet state, or when small difference in resonance frequencies of two coupled atoms reveals sharp spontaneous emission spectrum [19]. The Hamiltonian of dipole-dipole couplings $H_d$ is invariant under π-rotation by the operator $\pi_X = exp(i\pi S_X)$ with eigenvalues ±1. Therefore, all egenfunctions of $H_d$ can be rearranged to be eigenfunctions of $\pi_X$ and classified according to their "$\pi_X$-parity". These functions are no longer eigenfunctions of $S_Z$ but linear combinations of functions with ± $m$. $S_X$ has no matrix elements between the functions of different parity, so its evolution happens independently in the two subspaces. The small operator Δ breaks this symmetry, and the echo described in this paper allows seeing the slow evolution between the subspaces.

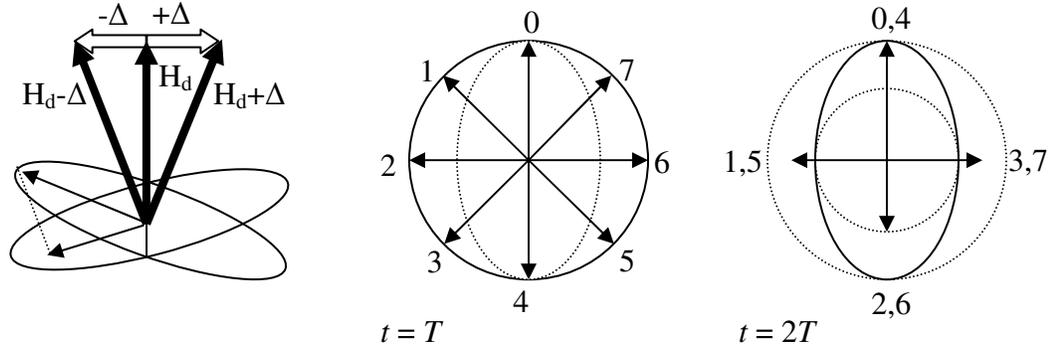

FIG. 3. *The echo formation. a) Projection of "old" operators on "new" operators; b) dephased components at t = T; c) partial echo at  t = 2T.*

## IV. EXAMPLES

All NMR experiments have been performed at ambient temperature (22°C) using Varian Unity/Inova 500 MHz NMR spectrometer. Solid samples, except naphthalene, have been dried in open NMR tubes at 110°C to eliminate possible contamination by adsorbed water.

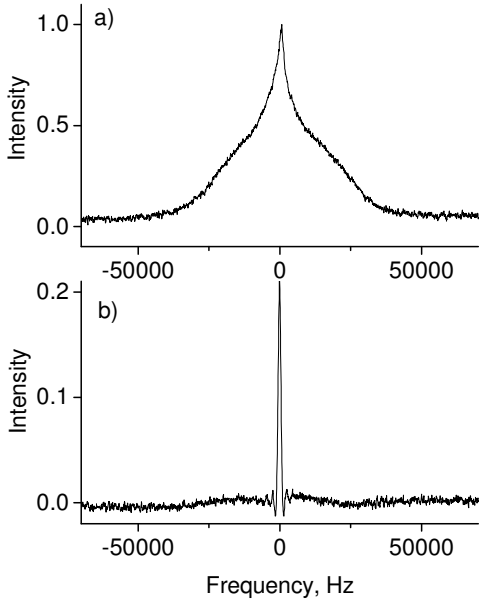

FIG. 4. *a) Proton NMR spectrum of glucose; b) the spectrum excited by 0.5 ms rectangular pulse with 30° flip angle.*

Fig. 4a shows the conventional proton NMR spectrum of glucose. The spectrum is broad, about 50 kHz. The distribution of isotropic chemical shifts is broad for this sample, so we cannot expect very narrow response signals. The spectrum excited by 0.5 ms rectangular pulse with 30° flip angle, followed by the refocusing $\pi$-pulse, is presented in Fig. 4b. Visible 2 kHz oscillations reflect the excitation spectrum of 0.5 ms rectangular pulse and suggest that the peak can be narrower for longer excitation pulses.

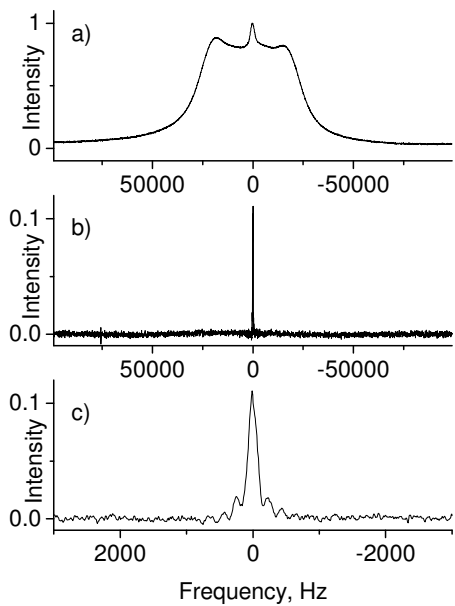

FIG. 5. *a) Proton NMR spectrum of naphthalene; b) and c) the spectrum excited by 5 ms rectangular pulse with 30° flip angle.*

Proton NMR spectrum of naphthalene is displayed in Fig. 5a. The result of excitation by 5 ms rectangular pulse with 30° flip angle is shown in Fig. 5b. Difference of isotropic chemical shifts for this sample is about 0.3 ppm (150 Hz), which is approximately equal to the limiting signal narrowing, which can be achieved for naphthalene at increased duration of the excitation pulse. The expanded signal in Fig. 5c shows expected 200 Hz oscillations (distorted, because the peak width approaches its limiting value). Naphthalene has very long $T_1$ relaxation time for protons, so $10^3$ s repetition time has been used. The spectra in Figs. 5a and 5bc are the results of 16 and 64 averages, respectively. It should be noted that the broad spectra in Figs. 4 and 5 are distorted, and presented only as illustrations. The spectrum in Fig. 4a has been recorded by using 10 µs solid-echo [20]. The spectrum in Fig. 5a is a single-pulse acquisition with 9 µs delay between the center of $\pi/2$ pulse and the first acquired point.

The example in Fig. 6 is shown because it does not fit the theoretical discussion above. Two non-equivalent protons of polybutadiene are clearly resolved in Fig. 6a. The spectrum looks like a spectrum in liquid with insufficiently fast molecular motions. Usual Hahn echo for this sample has a decay time 1.3 ms, which accounts for about 50% of the line widths in Fig. 6a. Normally, it would be viewed as the "true" homogeneous linewidth, limited by spin-lattice relaxation, resulting from random molecular motions. However, 50 ms rectangular pulse with 30° flip angle, followed by $\pi$-pulse, produces much narrower signal shown in Fig. 6b. Here each of spin species moves almost independently, coupled to another by weak residual dipolar coupling, so the effect of the $\pi$-pulse will produce only a small perturbation. If we compare the signal in Fig. 6b to the long-lived singlet state in ref. [17], we may guess that the selective excitation reveals some multi-spin states, which are immune to spin-lattice relaxation. Formally, replacement of $H_d$ by $H_d + F$, where $F$ is time-independent Hamiltonian of the lattice, makes the problem similar to the one we already discussed.

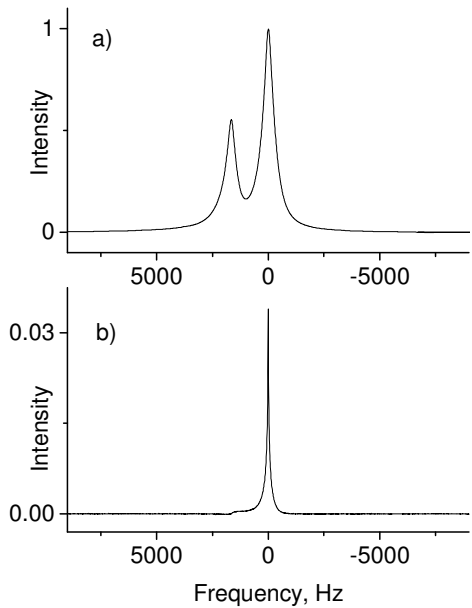

FIG. 6. *a) Proton NMR spectrum of polybutadiene; b) the spectrum excited by 50 ms rectangular pulse with 30° flip angle.*

## V. THE ECHO

The induction signal after the refocusing pulse reproduces the exact shape of the excitation pulse, including possible delays. Therefore, if an additional delay is introduced between the excitation and the refocusing pulses, the same delay will appear between the refocusing pulse and the echo signal. Fig. 7a shows the decay of the echo signal in adamantane, excited by 2 ms and 4 ms rectangular pulses, as a function of the echo time. Total flip angles of both pulses are the same, 30°. The echo time is the interval between the centers of the excitation pulse and the echo signal. The refocusing hard $\pi_X$-pulse is in the middle of this interval. The echo amplitude decays exponentially with 15 ms decay time. The actual echo amplitude for 2 ms pulse is twice the amplitude for the 4 ms pulse. In Fig. 7a the amplitudes are scaled for convenient comparison. The echo decay time is consistent with the smallest linewidth of the signal, about 50 Hz, which can be excited in adamantane at increased duration of the excitation pulse. This limit can be viewed as a new "homogeneous" linewidth. However, it does not reflect any true irreversible process in the system, but rather the method of the echo creation, which utilizes the difference of chemical shifts.

Fig. 7b shows the result for the Hahn echo sequence: $(\pi/2)_Y$-$T$-$\pi_X$, with 4 µs and 8 µs $(\pi/2)_Y$ and $\pi_X$ pulses, respectively. As a function of the echo time $2T$, the echo decays exponentially with the decay time 12.8 ms. The echo amplitude is plotted here as a fraction of the initial amplitude of the conventional free-induction decay signal excited by a single $(\pi/2)_Y$ pulse. The echo has duration, at half height, about 0.5 ms, which is about ten times longer than the conventional free-induction decay signal. The frequency of the echo signal coincides with the center of the NMR line for adamantane. The inverse of the echo width gives a range of frequency offsets from the NMR line center, where sharp selective signals can be excited. This range of offsets is, therefore, considerably narrower than the original linewidth of the spectrum.

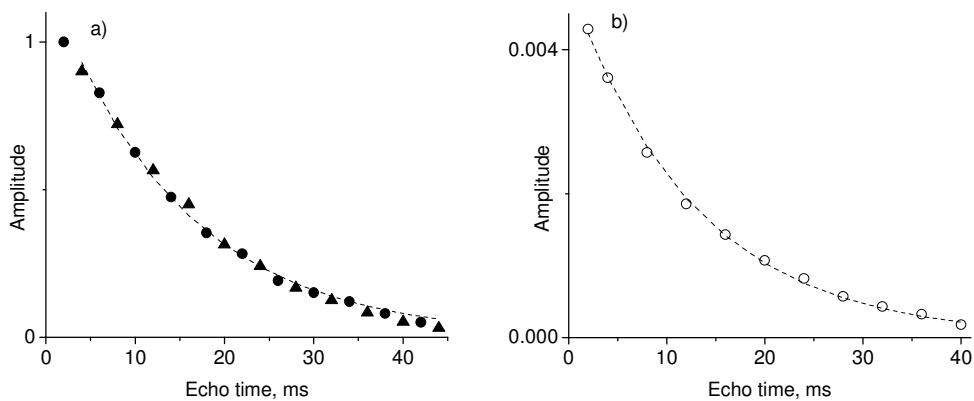

FIG. 7. *The echo decay in adamantane. a) The signal is excited by 2 ms (circles) and 4 ms (triangles) rectandular pulses with $30^o$ flip angle. The decay time of the fitting exponential curve is 15 ms. b) The echo is excited by hard pulses: $(\pi/2)_Y$-$T$-$\pi_X$, $2T$ is the echo time. The amplitude is in fractions of conventional free-induction signal after $(\pi/2)_Y$ pulse. Fitting exponential decay curve has 12.8 ms decay time.*

Sharp spectra in Figs. 2 and 4-6 are excited at the frequency of the soft excitation pulse. Therefore, their central frequencies do not carry direct spectroscopic information. The situation is different for non-selective Hahn echo sequence. In this case the echo signal appears at the central frequency of the conventional spectrum. Since the echo duration is longer than the free induction decay, the Fourier transform of the echo signal gives the spectrum with enhanced spectral resolution. Fig. 8a shows the conventional proton NMR spectrum of adamantane. The spectrum in Fig. 8b with 10-fold better resolution is the Fourier transform of the Hahn echo signal at 2 ms echo time.

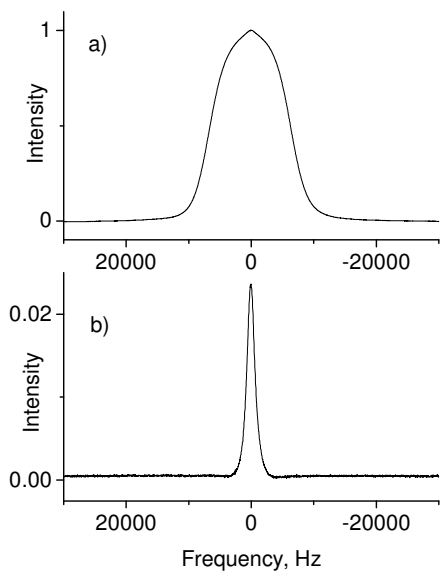

FIG. 8. *a) Proton NMR spectrum of adamantane: a) Fourier transform of free induction signal; b) a Fourier transform of the echo signal at $2T = 2$ ms echo time for the Hahn echo sequence $(\pi/2)_Y$-$T$-$\pi_X$-$T$.*

## VI. CONCLUSION

We described a new echo, based on symmetry breaking. If there is a symmetry operation, which leaves both the main Hamiltonian and the observable invariant, and there is also a perturbation, breaking this symmetry and non-commuting with the Hamiltonian, a change of the sign of this perturbation (or, more generally, just change) causes an echo. If this change happens at time $t = T$, after free evolution at times $0 < t < T$, the partial echo is formed at $t = 2T$. The amplitude of this echo is proportional, and the decay rate is inversely proportional to the magnitude of the symmetry-breaking perturbation. The echo can be also viewed as resulting from partial reversal of the dynamic evolution, and it seems that the echo may exist under very broad assumptions about the nature of this dynamic evolution. Matrix elements $A_{mn}$ of any operator $A$ in the basis, where the Hamiltonian is diagonal, evolve as $A_{mn}(t) = A_{mn} exp(i(E_m - E_m)t)$, where $E_m$ and $E_n$ are the energy levels of the Hamiltonian. The partial echo is created by mixing to the components $A_{mn}$, which evolve in the "forward" time direction, small fractions of the components $A_{nm} = A_{mn}*$, which evolve "backward" in time.

**References**

[1] E. Hahn, Phys. Rev. **80**, 580 (1950).

[2] B. N. Provotorov, Zh. Eksp. Teor. Fiz. **41**, 1582 (1961).

[3] M. Goldman, *Spin Temperature and Nuclear Magnetic Resonance in Solids*, (Clarendon, Oxford, 1970).

[4] E. B. Feldman, A. K. Khitrin A.K., Phys. Lett. A **206**, 128 (1995); A. K. Khitrin, E. B. Fel'dman E.B., Czech. J. Phys. **46, S4**, 2217 (1996).

[5] A. K. Khitrin, V. L. Ermakov, B. M. Fung, Chem. Phys. Lett. **360**, 161 (2002).

[6] A. K. Khitrin, V. L. Ermakov, B. M. Fung, J. Chem. Phys. **117**, 6903 (2002).

[7] A. K. Khitrin, V. L. Ermakov, 2002, http://arxiv.org/abs/quant-ph/0205040.

[8] A. K. Khitrin, V. L. Ermakov, B. M. Fung, Z. Naturforsch. **59a**, 209 (2004).

[9] V. Antochshuk, M.-J. Kim, A. K. Khitrin, J. Magn. Reson. **167**, 133 (2004).

[10] M.-J. Kim, A. K. Khitrin, Magn. Reson. Imaging **23**, 865 (2005).

[11] M.-J. Kim, K. Cardwell, A. K. Khitrin, J. Chem. Phys. **120**, 11327 (2004).

[12] R. Kubo, J. Phys. Soc. Japan, **12**, 570 (1957).

[13] W.-K. Rhim, A. Pines, J. S. Waugh, Phys. Rev. Lett. **25**, 218 (1970).

[14] W.-K. Rhim, A. Pines, J. S. Waugh, Phys. Rev. B **3**, 684 (1971).

[15] S. Ding and C. A. McDowell, J. Magn. Reson. *A* **111**, 212 (1994) .

[16] S. Ding and C. A. McDowell, J. Magn. Reson. *A* **117**, 171 (1995) .

[17] M. Carravetta, O. G. Johannessen, M. H. Levitt, Phys. Rev. Lett. 92, 153003 (2004).

[18] R. Sarkar, P. R. Vasos, G. Bodenhausen, J. Am. Chem. Soc. **129**, 328 (2007).

[19] J.-S. Lee, M. A. Rohrdanz, A. K. Khitrin, J. Phys. B: At. Mol. Opt. Phys. **41**, 045504 (2008).

[20] J. G. Powles, P. Mansfield, Phys. Lett. **2**, 58 (1962).

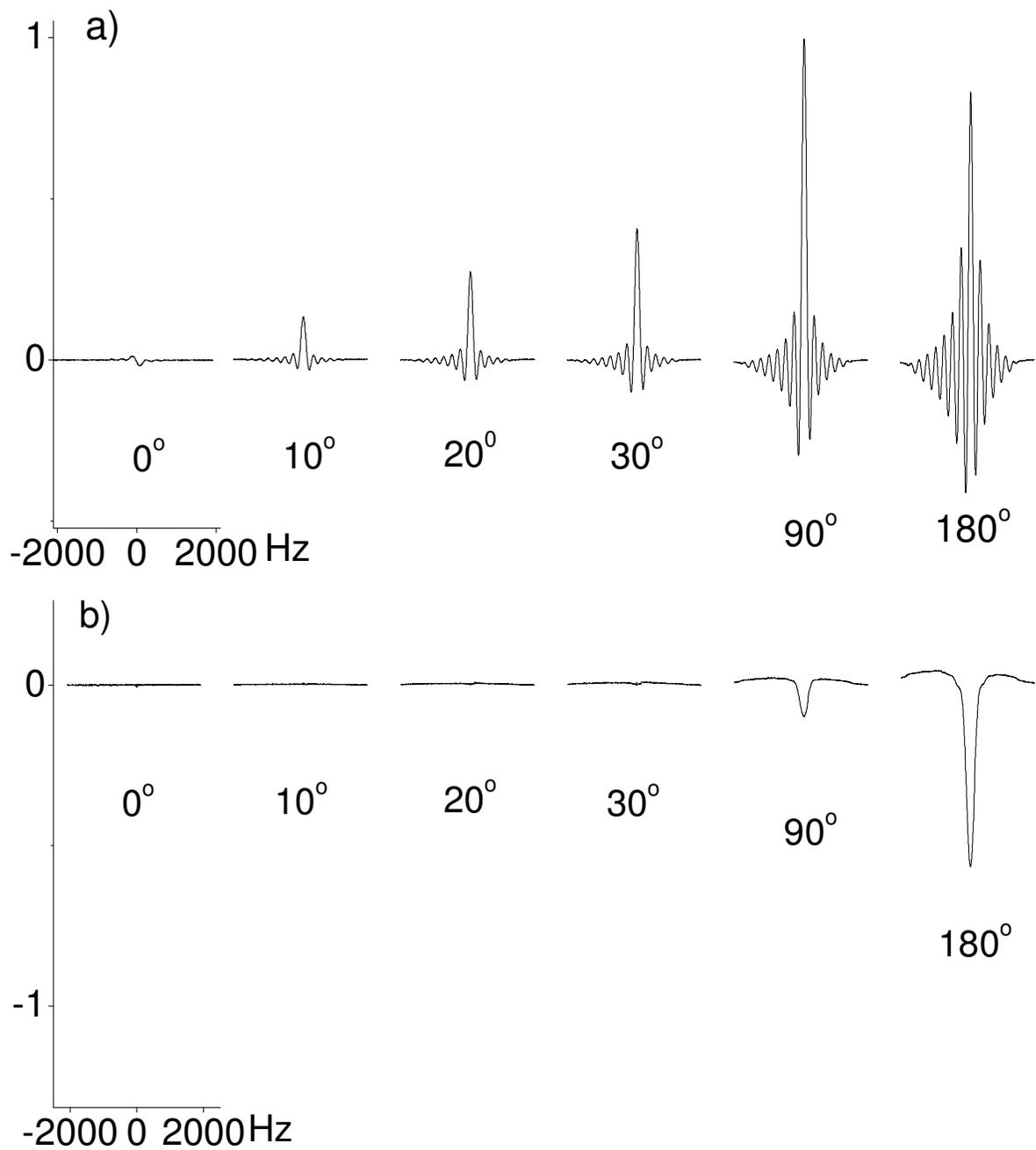

FIG. 1

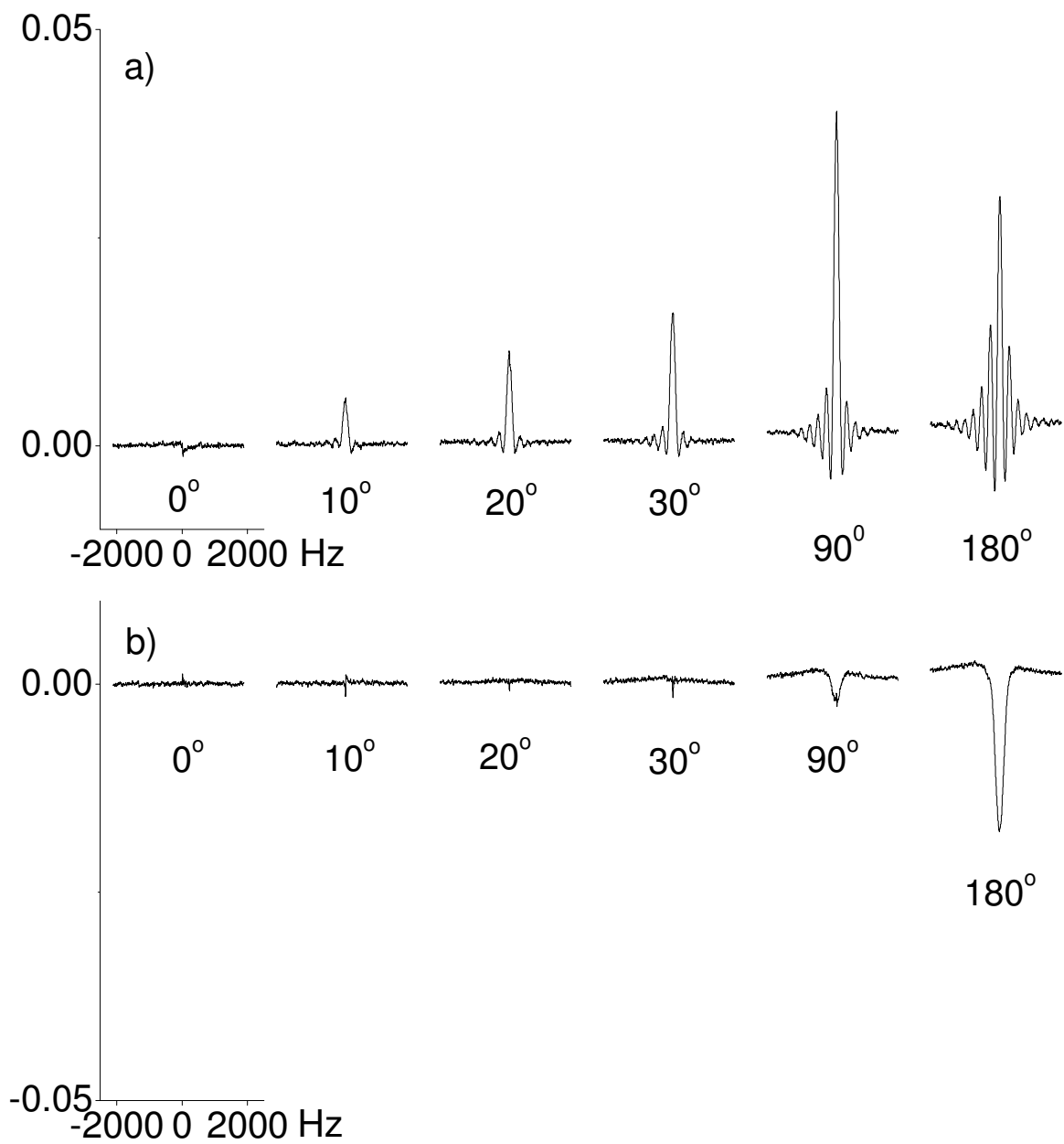

FIG. 2

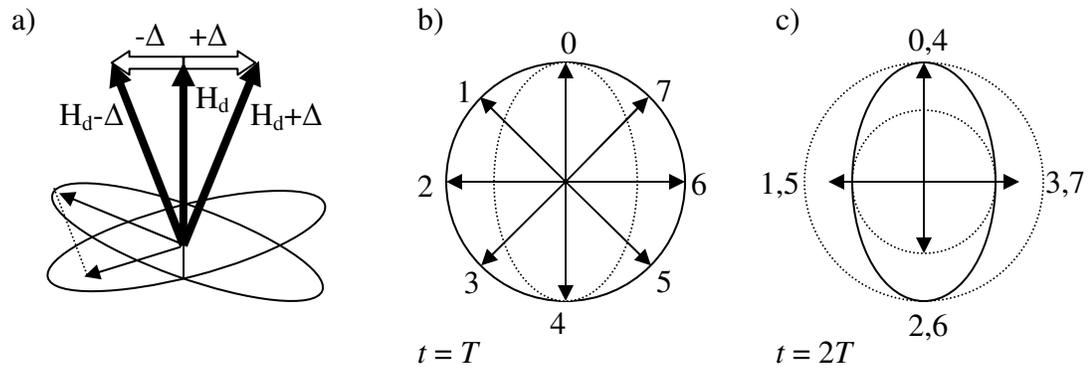

a)

$-\Delta$ $+\Delta$

$H_d-\Delta$ $H_d$ $H_d+\Delta$

b)

$t = T$

c)

$t = 2T$

FIG. 3

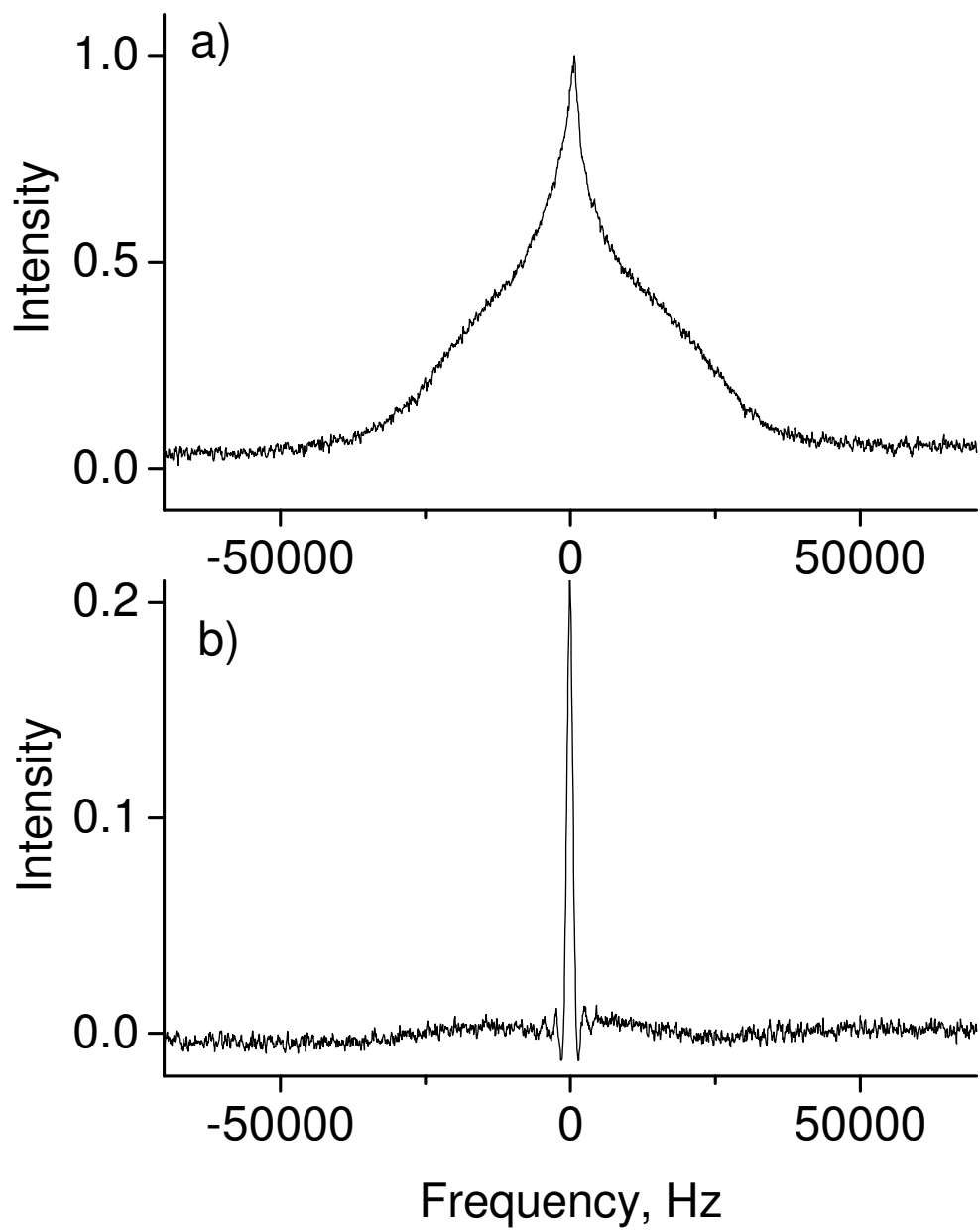

FIG. 4

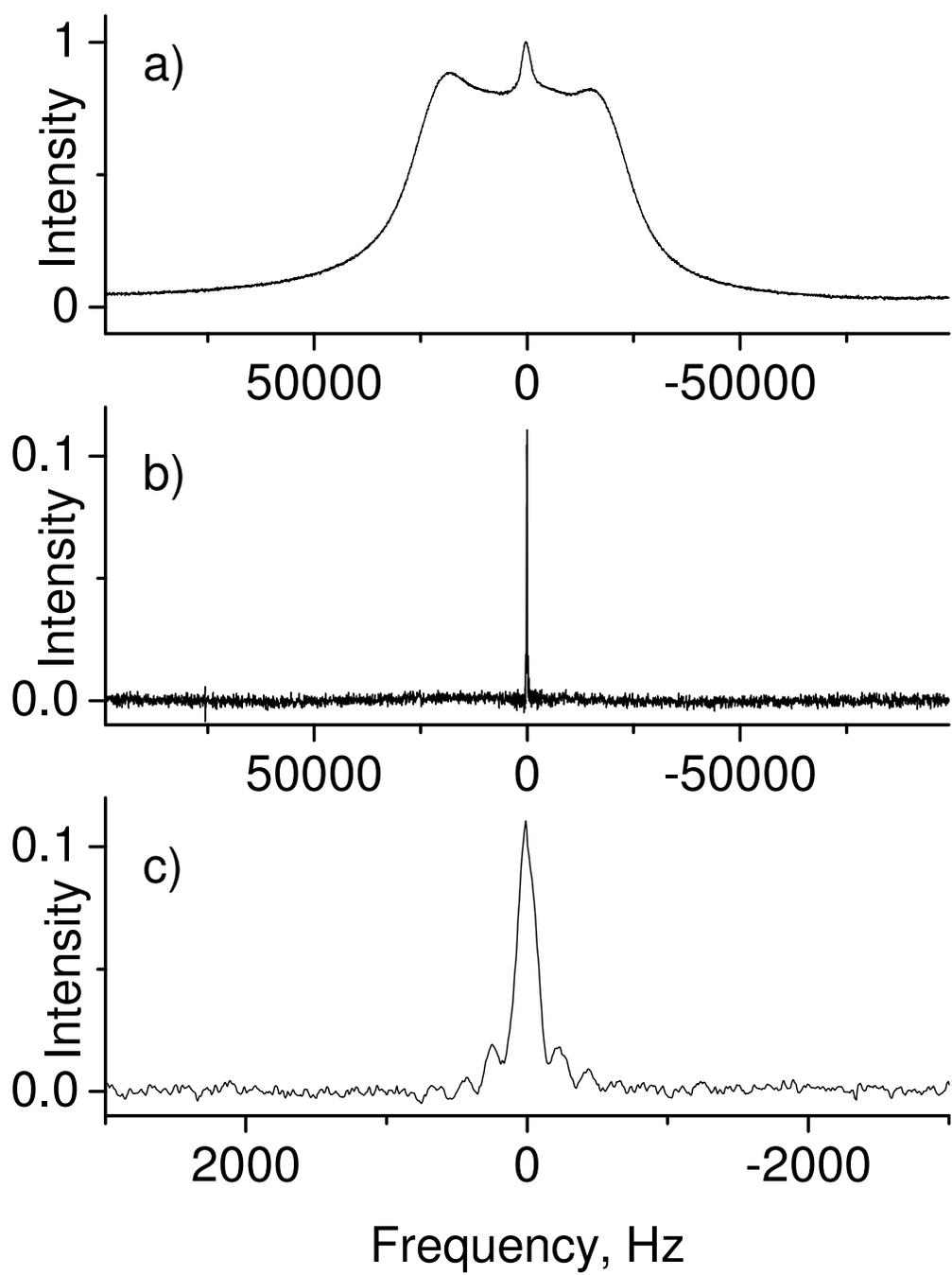

FIG. 5

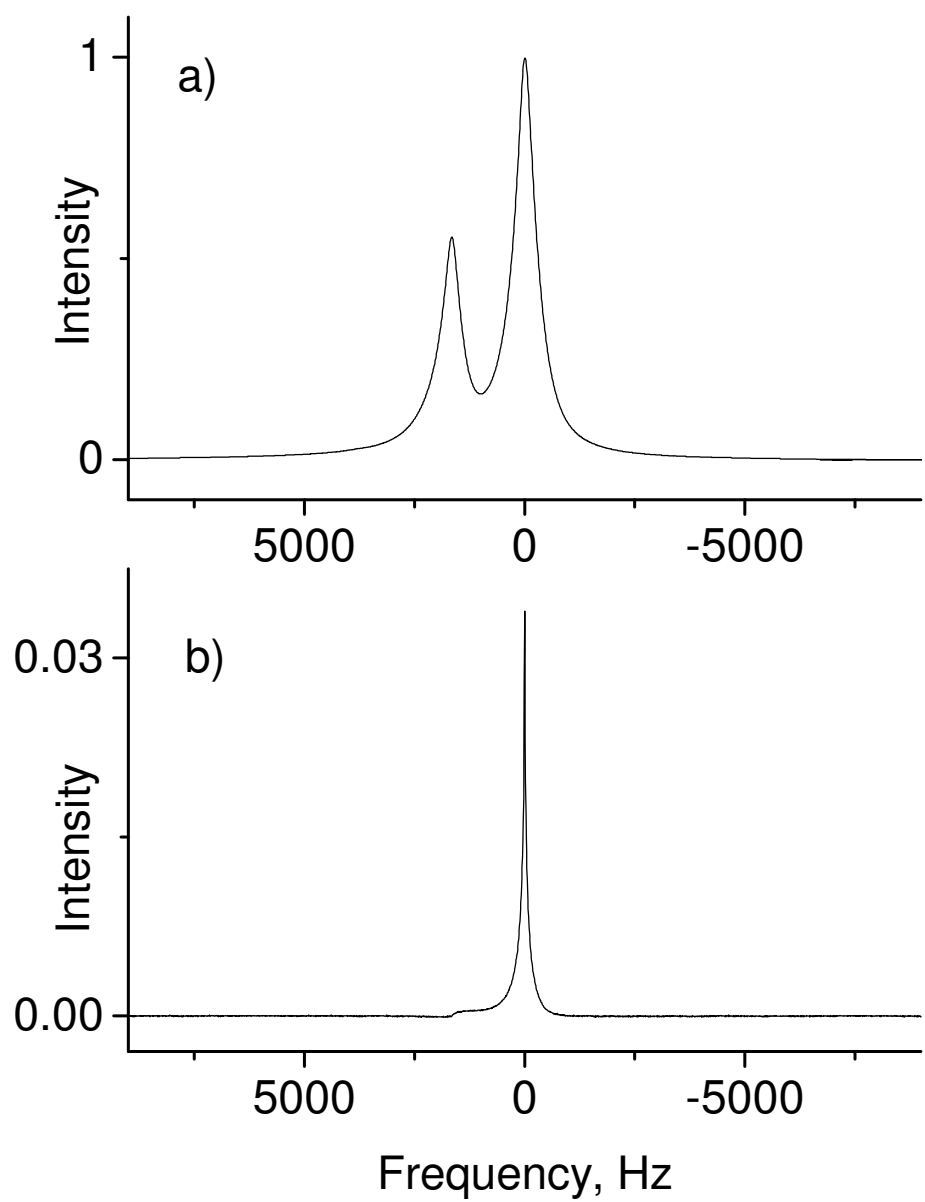

FIG. 6

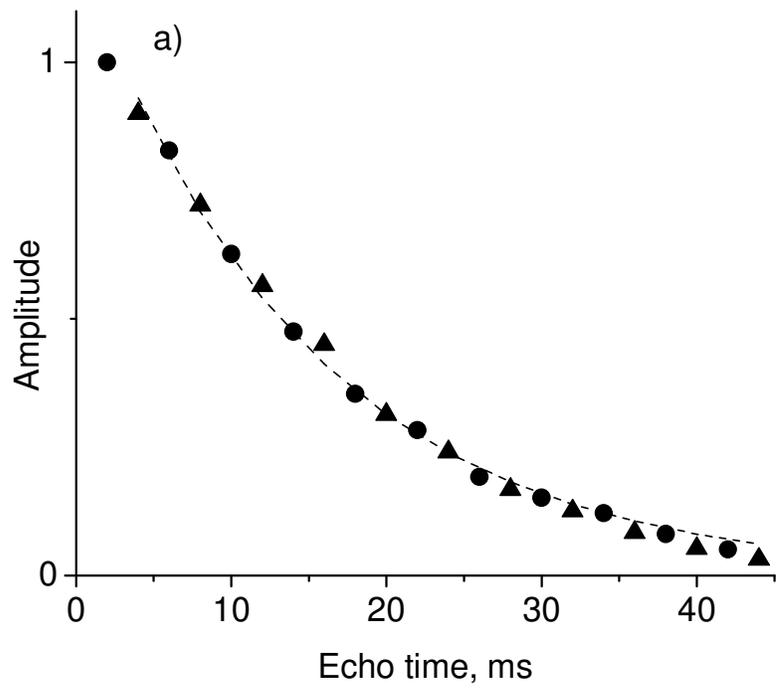

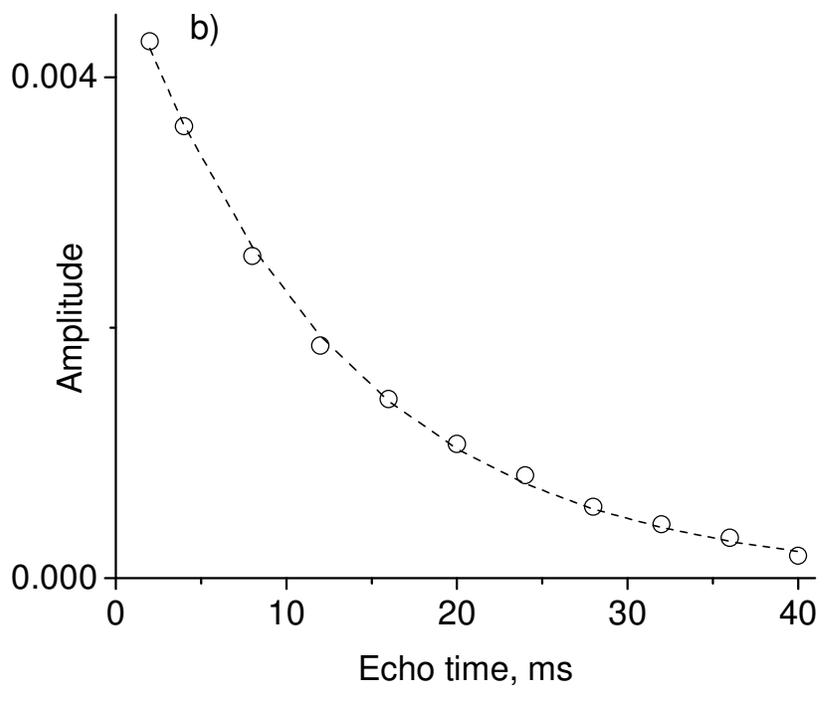

FIG. 7

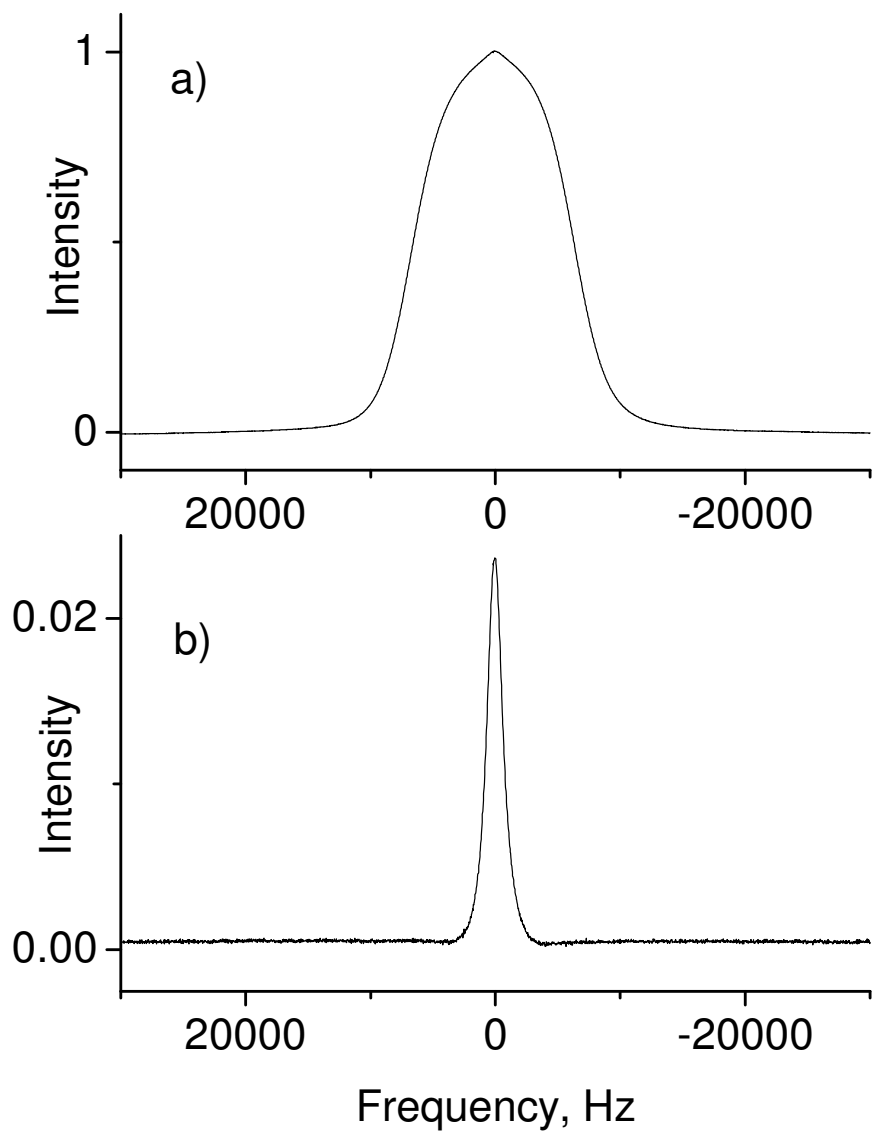

FIG. 8